\documentclass[manuscript]{acmart}
\AtBeginDocument{%
  }

\usepackage{subcaption}
\usepackage{xcolor}
\usepackage[T1]{fontenc}
\usepackage{hyperref}

\begin{document}

\title{Lost in Transcription: Subtitle Errors in Automatic Speech Recognition Reduce Speaker and Content Evaluations}


\author{Kowe Kadoma}
\email{kk696@cornell.edu}
\affiliation{%
  \institution{Cornell University}
  \city{Ithaca}
  \state{NY}
  \country{USA}
}

\author{Priyal Shrivastava}
\email{pshriva2@cs.cmu.edu}
\affiliation{%
  \institution{Carnegie Mellon University}
  \city{Pittsburgh}
  \state{Pennsylvania}
  \country{USA}
  }

\author{Mor Naaman}
\email{mor.naaman@cornell.edu}
\affiliation{%
  \institution{Cornell Tech}
  \city{New York}
  \state{New York}
  \country{USA}
}

\renewcommand{\shortauthors}{Kadoma et al.}
\newcommand{\minititle}[1]{\noindent
\textbf{#1}}

\begin{abstract}
Researchers have demonstrated that Automatic Speech Recognition (ASR) systems perform differently across demographic groups. In this work, we examined how subtitle errors affect evaluations of speakers and their content using a preregistered online experiment (N=207, U.S.-based crowdworkers). Participants watched speakers with various accents deliver a talk in which the subtitles were accurate or error-prone. Our results indicate that error-prone subtitles consistently reduce both speaker and content evaluations for all speakers. 
We did not see disparate impact between the accent groups, controlling for subtitle quality. Taken together, though, the findings of this short paper imply that speakers with accents for which ASR systems perform poorly are likely to be further penalized by viewers with lower evaluations.
\end{abstract}




\maketitle

\section{Introduction}
Automatic speech recognition (ASR) systems convert spoken language into text using signal processing and machine learning~\cite{lu2019automatic,synnaeve2020endtoendasrsupervisedsemisupervised}.
These systems now power increasingly ubiquitous technologies, from voice assistants like Apple Siri and Amazon Alexa to transcription services and automatic subtitles on platforms like YouTube, Zoom, and Microsoft Teams.
Automatic subtitles facilitate accessibility, enable cross-linguistic communication, and support users in noisy environments. 
Beyond convenience, they play a critical role in providing equitable access to information for individuals who are deaf or hard of hearing~\cite{kuhn-dhh-2024,arroyo-2024-subtitleQuality,desai-2025-subsMultiAccessability} as well as non-native speakers~\cite{shimogori-2010-captionsESL}.

Despite their promise, ASR systems are not error-free, and their accuracy is not distributed evenly across different users. 
ASR systems often exhibit biases, or systematic differences in performance,
with respect to race~\cite{koenecke_asr_2020,ezema-interfaces-2025,harrington-switch-2022}, gender~\cite{hutiri_bias_2022,Tatman2017EffectsOT,hwang-2019-gender}, and dialect~\cite{harris-etal-2024-modeling,Tatman2017EffectsOT}.
For example, researchers found that the embeddings (or word associations) in the pre-trained speech models of ASR systems often associated positive words with abled individuals over those with disabilities, White Americans over Black Americans, men over women, and U.S. accents over non-U.S. accents~\cite{slaughter-etal-2023-pre}.

These performance differences have real consequences for the individuals whose speech is being analyzed~\cite{harrington-switch-2022,wenzel_voice_2023,li-research-2025,kim2024same}.
Wenzel et. al~\cite{wenzel_voice_2023} found that when ASR errors occur, Black speakers experience significantly higher levels of self-consciousness and lower self-esteem compared to White speakers. 
In the context of automatic subtitles for Q\&A sessions in conferences, Li et. al~\cite{li-research-2025} demonstrated that non-native English speakers experience frustration and confusion when subtitle errors occur. Furthermore, non-native English speakers were concerned about the potential social stigma that comes with using ASR tools. Participants in the study expressed concern that using an ASR tool could lead the audience to perceive them as less competent, hinting at the importance of how the audience perceives the speaker~\cite{li-research-2025}.
If an ASR tool can affect how the audience perceives the speaker, then individuals whose accents produce more ASR errors may face a compounded disadvantage. Not only are their talks more difficult to understand and less enjoyable due to subtitle errors~\cite{chan2019comparing,Szarkowska10102024}, but their clarity and competence are also undermined~\cite{li-research-2025}, which could lead to further negative consequences in professional and social settings.

In this short paper, we examine the relationship between subtitle quality and evaluations of the speaker and the content through a carefully designed online experiment.
Our work builds on Kim et al.~\cite{kim2024same}, who compared subtitle errors across native and non-native speakers. Their design, however, did not control for confounders like accent with speaker identity and appearance. 
In this work, we overcome this limitation by using generative AI to manipulate the accent while holding all other speaker characteristics constant.
In our experiment, participants watched videos of broad-audience talks and evaluated the speaker and the content of each talk.
Participants watched two videos each, which were randomly selected from a set of four videos. 
Our experimental manipulation could present each of these videos alongside accurate or error-prone subtitles (the first manipulated factor) with the speaker using a Standard American English (SAE) accent or a non-Standard American English (non-SAE) accent (the second factor). 
A non-SAE accent could refer to regional English varieties within the United States or non-standard or globally regional accents; however, in this work, we examine an English variety spoken in India.

Our results show that subtitle errors negatively impact the evaluation of the speaker and the content. 
Given the widespread deployment of ASR systems across professional, educational, and social platforms, such errors may disadvantage a large number of speakers. The negative evaluations could be particularly concerning for individuals whose accents trigger higher error rates; however, we did not see direct evidence that the speaker's accent further impacts their evaluations when subtitle errors occur. 
In addition to our empirical findings, our work offers a key methodological contribution in the form of a novel experimental design that controls for the speakers' appearance.
In total, we provide more robust evidence of the downstream effects of subtitle errors, highlighting the need to consider the potential adverse impacts of these technologies.

\section{Background and Hypotheses}
Subtitles are widely used in asynchronous settings, such as movies, TV, and online platforms like YouTube, as well as in live, synchronous settings, such as video conferencing software like Zoom. 
While subtitles were originally introduced in film and television to support deaf and hard-of-hearing audiences~\cite{gernsbacher2015video}, research has shown that hearing viewers also benefit from subtitles through improved comprehension and engagement~\cite{chan2019comparing,taylor2005perceived,kruger-2013-subCognitiveLoad}. 
These benefits are often explained by the dual coding theory, which posits that people remember information more effectively when it is presented through both audio and visual channels~\cite{clark1991dual}.
The effectiveness of subtitles, however, hinges on their quality. 
Many factors impact subtitle quality, including accuracy, display size, and speed~\cite{chan2019comparing,kushalnagar-2015-subSize,ROMEROFRESCO201656}.
Researchers, user associations, and regulators have extensively studied subtitle quality, specifically subtitle accuracy, finding that inaccurate subtitles reduce the audience's comprehension and engagement~\cite{apone2010caption,english2014report,waes2013live}. 
Inaccurate subtitles do more than limit audience understanding---they can potentially negatively impact the speaker. 
ASR-generated subtitles are often less accurate than human-generated subtitles~\cite{romero2025fit,romero2023accuracy,chan2019comparing}, meaning that speakers in video lectures, interviews, or live calls may be unfairly penalized by errors outside their control. 

ASR systems are often less accurate due to biases that affect the system's performance~\cite{baumann-2023-biases,friedman-bias-1996}.
Biases occur when the model produces different outcomes for individuals belonging to different subgroups (e.g., gender or race) despite having the same ground truth (or human-verified scores).
Many biases are seen as byproducts of the models' construction and, in particular, the training data used~\cite{shelby_taxonomy_2023,weidinger-risks-2022}.
It is important to note that while training data is often seen as a source of bias, it is not the only source of bias in the machine learning life cycle. For example, bias could emerge when the methods used to compute features used by the ML models are themselves biased or when the features subtly encode protected
features like gender or race~\cite{harini-sources-2021,booth-bias-2021}. 
Several studies have examined biases in ASR systems across several demographic groups like race~\cite{koenecke_asr_2020,ezema-interfaces-2025,mikel-asr-2022}, gender~\cite{hutiri_bias_2022}, and dialect~\cite{harris-etal-2024-modeling,lima-vaBias-2019,Tatman2017EffectsOT,hutiri_bias_2022}.
In one of the most notable of these studies, Koenecke et. al~\cite{koenecke_asr_2020} showed that commercial ASR systems have a significantly lower word error rate (WER) for White speakers than Black speakers, even when the speakers uttered identical phrases.

When ASR systems produce errors as a result of biases, users from marginalized backgrounds can disproportionately experience harms~\cite{shelby_taxonomy_2023,cunningham-etal-2024-understanding,wenzel-2024-reduction}. 
For example, such users may need to increase their effort to make an ASR system like a voice assistant work for them.
Harrington et al.~\cite{harrington-switch-2022} highlighted the difficulties
older Black speakers encountered in accessing health-related information via Google Home, requiring them to engage in a form of ``cultural code switching.'' 
Likewise, Cunningham et al.~\cite{cunningham-etal-2024-understanding}
noted that African American speakers often engage in a form of
``invisible labor'' by adjusting their natural speech patterns to make voice assistants work.
In addition to increased labor, users of marginalized backgrounds may experience emotional harm in the form of deep feelings of frustration, self-consciousness, and shame when ASR systems do not perform well~\cite{wenzel_voice_2023,mengesha_inclusive_2021}.

Most research on harms in ASR systems focuses on voice assistants, where failures typically occur in private settings. By contrast, errors in automatic captioning are public, introducing an audience whose judgments can amplify emotional harms to the speaker.
As Li et. al~\cite{li-research-2025} notes, non-native speakers, in particular, worry that relying on ASR captioning tools in settings like academic talks may reflect poorly on them. 
Without the added burden of an erroneous captioning tool, speakers with non-standard accents face other social pressures, as prior work shows that they are seen as less credible compared to speakers with standard accents~\cite{lev2010don,fuertes2012meta,lorenzoni2024does}.
Taken together, these findings suggest that subtitle errors can disproportionately disadvantage speakers with certain accents by exacerbating social biases and negatively impacting evaluations.
In this study, therefore, we examine the relationship between subtitle quality and the evaluations of speakers and content with the following questions:

\begin{itemize}
  \item[\textbf{RQ1}] How do subtitle errors impact the evaluation of the speaker and the content?
   \item[\textbf{RQ2}] How does the speaker's accent impact their evaluation when subtitle errors occur?
\end{itemize}

We hypothesize that:
\begin{itemize}
   \item H1: Speakers and their content will be negatively impacted when subtitle errors occur.
   \item H2: Speakers with non-Standard American English (non-SAE) accents will receive more negative evaluations than speakers with Standard American English (SAE) accents. 
\end{itemize}

\section{Methods}
\begin{figure*}[t]
\includegraphics[height=5cm]{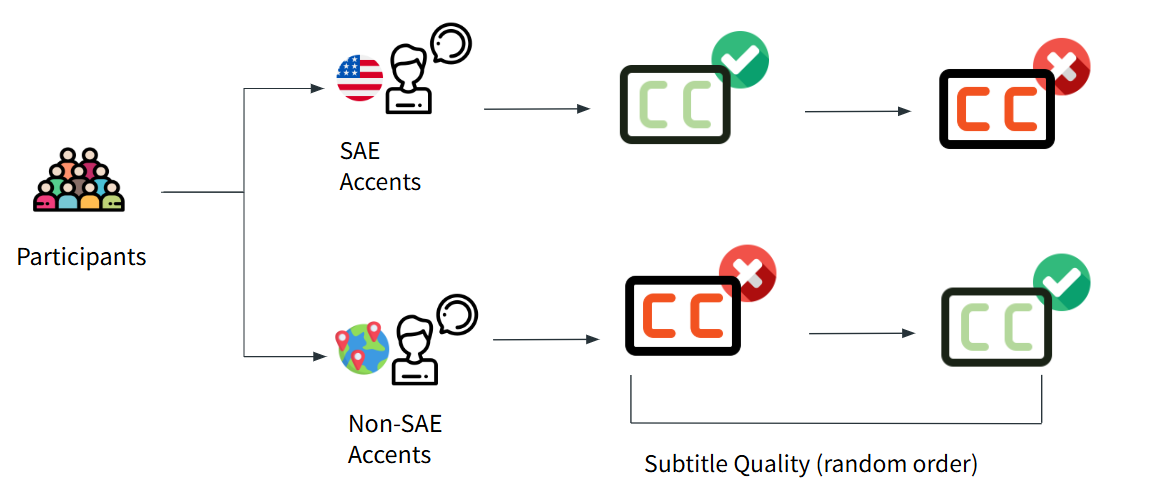}
\caption{An overview of the experimental procedure.}
\label{fig:experiment}
\end{figure*}

We conducted a mixed-factorial experiment to investigate how subtitle errors affect perceptions of the speakers and their content (RQ1) and how the speakers' accent influences these evaluations when subtitle errors occur (RQ2). An overview of the experimental design is illustrated in Figure~\ref{fig:experiment}.
As seen in the figure, participants were randomly assigned to one of two accent conditions (SAE or non-SAE) and watched two videos featuring speakers with the same accent. Each participant viewed one video with accurate subtitles and one with error-prone subtitles in a randomized order. After watching each video, participants evaluated the subtitle quality, the speaker and their delivery, and the content of the talk. Following the experiment, we collected demographic information.

The following sections detail the materials and procedure used in our experiment.
We first describe how we selected the videos (Section~\ref{sec:vids}) and created the subtitles (Section~\ref{sec:subtitles}), before describing our measurements (Section~\ref{sec:measures}) and participant recruitment (Section~\ref{sec:participant}). 
The research design was approved by Cornell's IRB and preregistered on OSF\footnote{\url{https://osf.io/9p6sb/?view_only=e773ccfebdb34b60b0b7d3c327682f3b}}.

\subsection{Video \& Audio Components} 
\label{sec:vids}
We selected publicly available TED (Technology, Entertainment, Design) Talks for our video stimuli since TED speakers address a broad audience, ensuring that the presentations are easy to understand.
For the validity of our accent manipulation, it is essential that the speaker is perceived as someone who could plausibly speak with either an SAE accent or a non-SAE accent. South Asian men were an appropriate choice for this purpose, as their representation in U.S. tech and business occupations is higher than their share of the U.S. workforce~\cite{varma2010india,nsf2024,immigration2022}. Additionally, English is one of India's official languages and widely used in educational and business settings~\cite{chandras2020multilingualismin,sridhar1996language}. As a result, without knowing where the speaker grew up, the speaker could plausibly be perceived as having either an accent heavily influenced by one of the regional languages of India or a more Americanized accent.

To further ensure ecological validity and avoid prior exposure, we selected regional TED Talks from India with fewer than 200,000 views and at least 5 years old.
We identified four talks in which the speakers (all men, likely in their 20s or 30s, and of South Asian appearance) discussed their entrepreneurial journeys.
The videos ranged from 13 to 21 minutes, and we extracted a 1-minute snippet from the middle of each talk. 
This duration aligns with prior research showing that trait impressions are formed within milliseconds of hearing a voice~\cite{mileva2023trait}, and observations beyond one minute yield diminishing returns for forming accurate behavioral impressions~\cite{cullen2017thin,murphy2021capturing}.

Because we are interested in how the accent affects the perception of the speaker, we needed to have two versions of each video---one with an SAE accent and one with a non-SAE accent. 
To ensure similar audio quality between the two accents, the audio in all of the videos was AI-generated using ElevenLabs, an AI voice-generation and manipulation service. 
Prior work has shown that compared to the voices from Speechify, another popular AI voice service, the voices from ElevenLabs are seen as more realistic with characteristics like stutters, mistakes, and breaths~\cite{michel_voice_2025}. 

We used ElevenLabs' English V2 model to produce two new audio tracks with different speaker accents for each video, one for each accent type. 
The model produced an audio speech track that closely matched the original audio timing, ensuring that the new audio could be aligned with the original video without discrepancy. 
We processed each video using ElevenLab's voice changer, using middle-aged, male voices, in a conversational and narrative/story style.
For the non-SAE condition, while the original voice already had the target accent, we still created a new voice to ensure that the video would have a high-quality recording without background noise or echoes while retaining an accent similar to the original. 
We used the voices of ElevenLabs' AI ``Hindi speakers'' with the default stability, similarity, and style exaggeration settings for the voice changer, meaning that the new voice closely resembles the original voice and retains or somewhat amplifies its characteristics.
For the SAE condition, we repeated a similar procedure with middle-aged, male ``American English'' AI voices. 
To ensure the American accent would be dominant in the generated voice, we set the similarity and style exaggeration to zero.
We then created two versions of each of our videos, replacing the original audio with the newly generated audio tracks. We performed several evaluations and validations for the audio quality and the audio-video synchronization, which can be found in Appendix~\ref{appendix:vids}.

\subsection{Subtitles}
\label{sec:subtitles}
We created accurate and error-prone subtitles for each video.
We used a human annotator (one of the authors) to create accurate (ground truth) subtitles.
To obtain realistic yet significant error-prone subtitles, we compared the transcriptions of our audio-processed videos across several popular platforms: Whisper, Google Meet, Zoom, YouTube, Otter, Apple, and Adobe.
To this end, we uploaded the videos with both SAE and non-SAE audio tracks to these services and evaluated each service's performance on each audio file using the word error rate (WER). 
WER is a standard measure of discrepancy between the human transcription (our ground truth) and the machine transcription. WER is defined as the sum of substitutions, deletions, and insertions between the machine and ground truth transcriptions, divided by the number of words in the ground truth. A higher WER means a more error-prone transcription~\cite{koenecke_asr_2020}. 

In our test, Google Meet was the most error-prone service with an average WER of 0.31.  
For each video, we chose the Google Meet transcription with the higher WER between the two accent conditions. 
As a result, our subtitles were a realistic example of errors that \textit{could} be created for a similar video by an actual system. 
Since our input to these models was relatively clear audio without background noise or other environment challenges, we believe the error rate could be well within the bounds of automatically generated transcriptions in the wild. We provide more details on the WER and examples of the types of errors in the Appendix, Tables~\ref {tab:wer_results} and~\ref {table:subtitle_examples}.

\subsection{Measures}
\label{sec:measures}
The direct measures we asked participants to evaluate included variables related to the quality of the subtitles (as a control and manipulation check), the speaker's delivery, and the content of the talk.

\minititle{Subtitle Quality.} 
We draw on speech and translation literature to develop comprehensive subtitle quality measures. 
Subtitle quality assessment requires examining multiple dimensions beyond accuracy, like timing and user experience~\cite{wilken-etal-2022-suber,arroyo-2024-subtitleQuality}.
The measures we used derive from established frameworks that assess subtitle effectiveness across functional, technical, and readability parameters~\cite{pedersen2017far}.
We focus on technical performance with the statements \textit{the subtitles were accurate}, and \textit{the subtitles were timely \& synchronized}.
We also explore the impact of subtitles on cognitive processing, as prior work has shown that subtitle presence can influence cognitive load~\cite{li2024automatic,kruger-2013-subCognitiveLoad}. We capture both the potential cognitive benefits and the costs of subtitle processing with the following statements: \textit{the subtitles improved my understanding of what was said}, and \textit{the subtitles were distracting}.
The response options for all the statements ranged from \textit{strongly disagree} (1) to \textit{strongly agree} (5).
For our analysis, we perform a row-wise average across the responses to create the \textit{subtitle index} measure, which captures the overall subtitle quality.

\minititle{Speaker's Delivery.}
We draw from communication and behavioral science literature, specifically public speaking and audience engagement, to capture how the audience perceives the speaker.
We used survey items from Liu et. al~\cite{liu_psfes_2024} and Curtis et. al~\cite{curtis_speaking_2015} which included:
\textit{the speaker articulated and pronounced clearly} and \textit{speaker seemed knowledgeable about the topic}.
Similar to our subtitle quality measure, the response options ranged from \textit{strongly disagree} (1) to \textit{strongly agree} (5).
For our analysis, we calculate a row-wise average of these measures to create the \textit{speaker index} for our overall speaker evaluation measure.

\minititle{Content Quality.}
While the speaker measures capture \textit{how} the speaker conveyed ideas, the content quality measures capture the evaluation of the content of the presentation.
We modified the statements from Liu et. al~\cite{liu_psfes_2024} to omit unnecessary details about the speech opening or conclusion (e.g., the thesis statement is clearly presented).
We measure content with the following statements: \textit{the main points were clear} and \textit{the message felt genuine}. 
The response options were the same as those for the content questions.
Similarly, we create a \textit{content index} by performing a row-wise average across the three statements to be used in our analysis.

\subsection{Participant Recruitment and Checks}
\label{sec:participant}
We recruited 250 participants for our experiment using a standard U.S. sample of English-speaking adults on the crowdsourcing service Prolific.
To determine an appropriate sample size, we hypothesized a medium effect size using an ANOVA repeated measures, within-between interaction, aiming for 80\% power based on a pilot experiment.

To ensure that participants watched the videos with the audio on, we added an attention check question after each video. One attention check asked whether participants heard a beep during the video, whereas the other asked about background noise during the talk.
To ensure high-quality responses for our analysis, we excluded participants who failed the attention checks (19 participants) or correctly guessed the purpose of the experiment (17 participants). We removed an additional seven participants due to a bug that prevented them from viewing one of the videos.
In total, we had usable data from 207 participants.
Overall, 33\% of participants were between 45-54 years old, 77\% identified as White, and 41\% received a bachelor's degree. 
More details about the participants can be found in Appendix~\ref{appendix:dems}.

\section{Results}
Our results show the impact of ASR-generated subtitle quality on speaker and content evaluations and hint at potential differences between speakers with different accents.
Our analysis shows that subtitle errors negatively impact both the speaker and the content evaluations (RQ1).
At the same time, the analysis shows that the speaker's accent does not impact how they are perceived when subtitle errors occur (RQ2).

\begin{figure*}[h]
    \centering
    \begin{subfigure}[h]{0.45\textwidth}
        \centering
        \includegraphics[height=4.5cm]{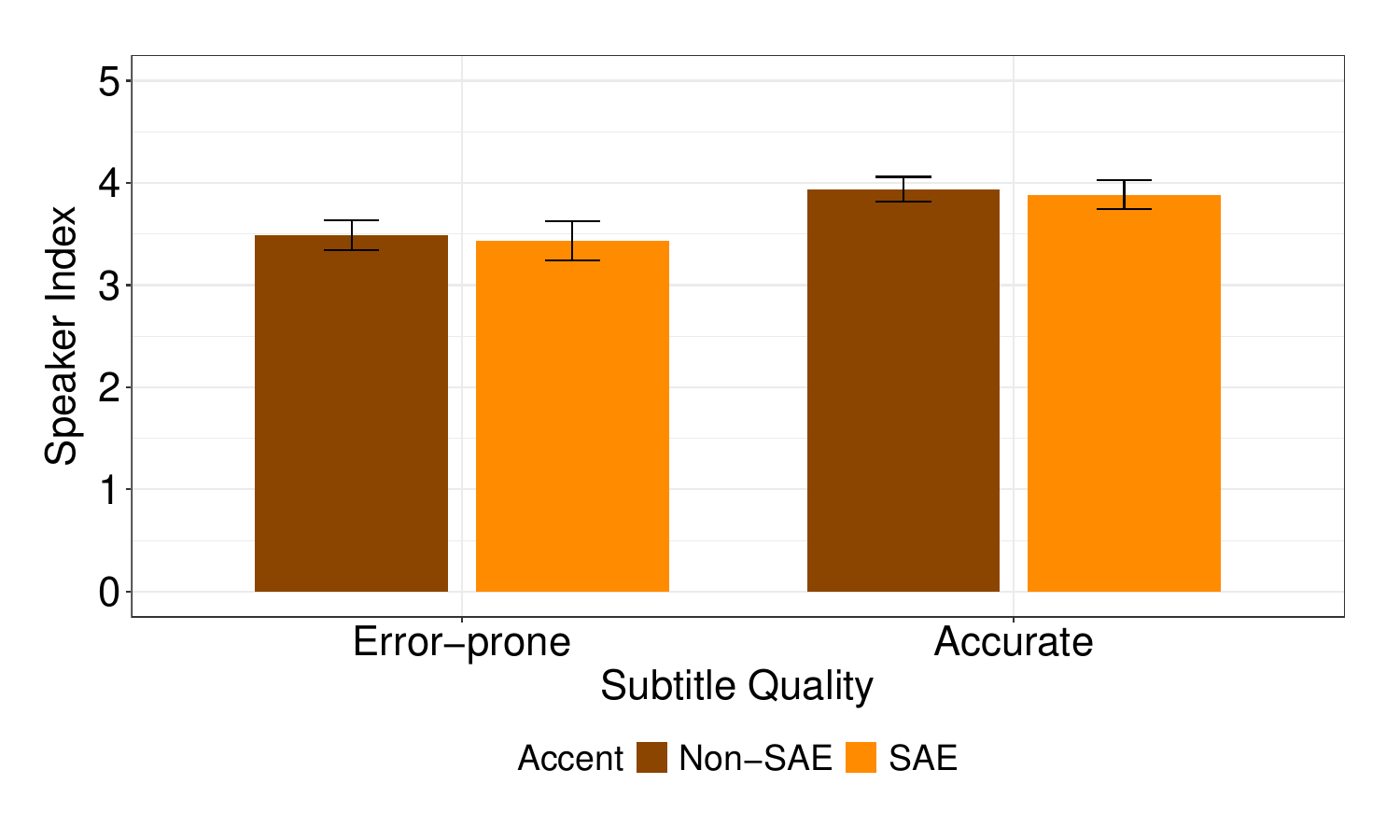}
        \caption{Accurate subtitles positively influence speaker evaluations.}
        \label{fig:speaker_index}
    \end{subfigure} 
    \hspace{10pt}
    \begin{subfigure}[h]{0.45\textwidth}
        \centering
       \includegraphics[height=4.5cm]{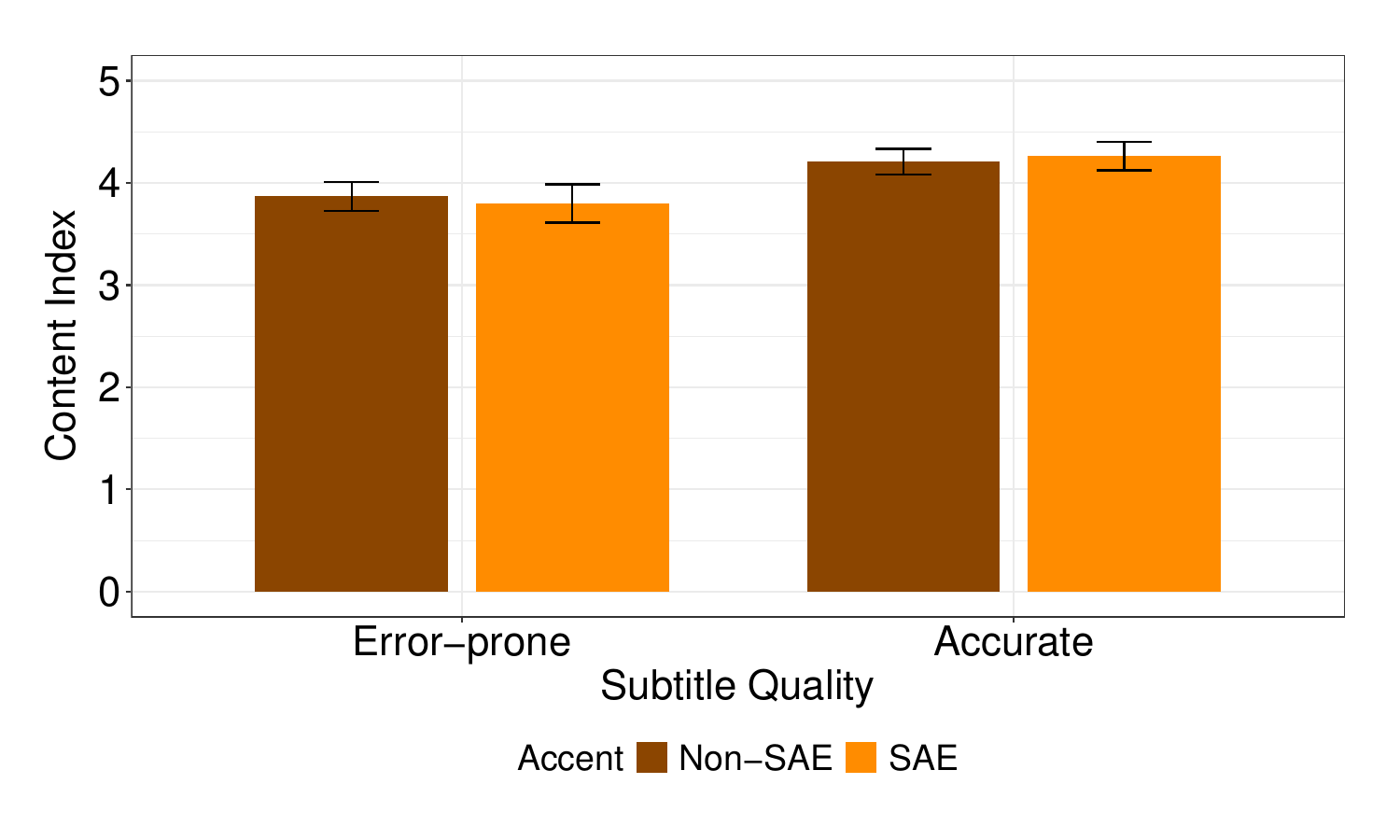}
       \caption{Accurate subtitles positively influence content evaluations.}
       \label{fig:content_index}
    \end{subfigure}
    \caption{The impact of subtitle quality on the speaker and content evaluations.}
    \label{fig:subtitle_effects}
\end{figure*}

Figure~\ref{fig:subtitle_effects} shows the mean, with a 95\% confidence interval, for our overall speaker evaluation measure (Fig.~\ref{fig:speaker_index}) and the overall content evaluation measure (Fig.~\ref{fig:content_index}) based on the 2x2 experimental manipulations of subtitle quality (x-axis) and speaker accent, with the SAE accent (in orange) and the non-SAE accent (in brown).
We first examine the impact of the subtitles on the speaker evaluations, as seen in Figure~\ref{fig:speaker_index}. The two columns on the left of the figure compare the evaluation scores given to speakers with error-prone subtitles (M\textsubscript{non-SAE}=3.48, M\textsubscript{SAE}=3.43).
These scores are lower than those of the same speakers with accurate subtitles on the right of the figure (M\textsubscript{non-SAE}=3.98, M\textsubscript{SAE}=3.88). 
From the figure, it appears as if there were no differences between speakers with SAE accents and speakers with non-SAE accents for a given subtitle quality.
To statistically test whether the observations in the figure are significant, we used a linear mixed model with subtitle quality and accent as fixed effects; we included participants as a random effect to account for the repeated measures design and individual variability.
Our analysis confirmed that subtitle quality has a significant main effect on speaker evaluations ($\beta = 0.451$, SE $= 0.082$, $z= 5.491$, $p<0.001$, 95\% CI [$0.290$, $0.612$]), but accent did not have a significant main effect ($\beta = -0.056$, SE $= 1.08$, $z= -0.519$, $p=0.604, n.s.$, 95\% CI [$-0.267$, $0.155$]). There were no interaction effects ($\beta = 0.001$, SE $= 0.122$, $z= 0.007$, $p=0.995, n.s.$, 95\% CI [$-0.238$, $0.240$]).
These results support H1, showing that the speakers receive better evaluations when there are accurate subtitles.

While the figure uses the averaged speaker evaluation measures, we performed a similar analysis and observed a similar trend for each of the individual speaker evaluation measures. The speaker is seen as clearer ($p<0.001$) and more knowledgeable ($p<0.001$) in the accurate subtitles condition.
The accent did not affect how clear ($p=0.981, n.s.$) or knowledgeable  ($p=0.426, n.s.$) the speaker appeared. 
There were no interaction effects between subtitle quality and accent (clarity: $p=0.503, n.s.$; knowledgeable: $p=0.475, n.s.$).

As Figure~\ref{fig:content_index} shows, the content evaluations are also higher when there are accurate subtitles (M\textsubscript{non-SAE}=4.21, M\textsubscript{SAE}=4.26) than when there are error-prone subtitles (M\textsubscript{non-SAE}=3.87, M\textsubscript{SAE}=3.80).
In line with our previous analysis, we employed a mixed-effects model, treating subtitle quality and accent as fixed effects and participant as a random effect.
The model confirmed that subtitles have a significant effect on how the content is perceived ($\beta = 0.341$, SE $= 0.078$, $z= 4.371$, $p < 0.001$, 95\% CI [$0.188$, $0.493$]), but the accent does not ($\beta = -0.069$, SE $= 0.106$, $z= -0.656$, $p = 0.512, n.s.$, 95\% CI [$-0.277$, $0.138$]).
There were no significant interaction effects between subtitle quality and accent ($\beta = 0.122$, SE $= 0.116$, $z= 1.055$, $p = 0.291, n.s.$, 95\% CI [$-0.105$, $0.349$]).
We see the same trend for our individual content measures, asking whether the content is clear and whether it is genuine.

\section{Discussion}
Taken together, our results provide some context on the impact of ASR-generated subtitle errors on speaker evaluations and who is most likely to be affected by them. 
Our study builds on Kim~\cite{kim2024same} by manipulating the accent within the same video stimuli. By doing so, we eliminate confounds inherent in comparing different speakers (e.g., facial features and gestures).
In this discussion, we address the question of why the ASR-generated errors reflect poorly on the speaker and content and under what circumstances the speaker's accent may impact their evaluation. 
Lastly, we discuss design implications and highlight the challenges and the importance of building more inclusive ASR systems.

We found that the speakers and the content of the talk were negatively impacted by subtitle errors (supporting Hypothesis~1), in contrast to Kim et al.'s earlier work showing that errors in ASR-based subtitles do not affect the perception of the speaker or their content~\cite{kim2024same}.
Whereas the earlier study only controlled for the speakers' accent and video's content, we controlled for the speakers' appearance, accent, and delivery style in this work.
By minimizing variability in speaker-related attributes, our design made subtitle quality the primary source of difference between conditions, which may have enabled our experiment to detect this effect.
The impact of subtitle errors on the speaker and content evaluations could be due to the fact that when subtitle errors occur, it can be more cognitively demanding for the audience and harm their comprehension and engagement~\cite{chan2019comparing,kruger-2013-subCognitiveLoad,taylor2005perceived}.

Although we found that speakers were negatively evaluated when subtitle errors occurred, our analysis revealed that speakers with non-SAE accents and speakers with SAE accents were equally penalized (leading us to reject Hypothesis~2). 
This finding is somewhat surprising, given prior work suggesting that social biases often lead to lower evaluations of speakers with non-standard accents, deeming what they say as less credible~\cite{lev2010don,fuertes2012meta,lorenzoni2024does}. 
Furthermore, prior work in ASR and accent evaluation did expose differences between non-standard (in that case, "non-native") and standard ("native") accents~\cite{kim2024same}.
One possible explanation for the lack of an effect in our experiment lies in our accent manipulation. To maintain the speaker's image and the talk's content, we relied on an AI-generated American voice overlaid on audio originally produced with an Indian accent. As a result, the final voice may not have sounded fully American. 
Consistent with this argument, our pre-test (see Appendix~\ref{appendix:vids}) found a statistically significant difference between the accents, but the effect size was small.

The implications of our findings may be substantial, with ASR-generated subtitles used across a variety of consequential domains, including education, law, business organizations, and live broadcasting.
For example, many organizations are increasingly using ASR-based subtitling tools during video interviews~\cite{hickman2024automated}.
As a part of the hiring process, candidates will record themselves answering predefined interview questions without an interviewer present. The video will then be uploaded to a review system where the ASR system will create a transcript and subtitles for the video, which will later be evaluated by the hiring manager. 
Prior work found that the ASR systems for video interviews produced more error-prone subtitles and transcriptions for speakers with non-standard accents~\cite{hickman2024automated}; taking into account our finding that subtitle errors harm speaker evaluations overall, job applicants with non-standard accents could encounter additional challenges when finding a job. 
Even within the organization, these speakers could still encounter challenges.
Speakers with non-standard accents already tend to receive lower evaluations in workplace settings, even when their performance matches that of speakers with standard accents~\cite{spence2024your,gluszek2010way}.
When video-conferencing platforms such as Zoom or Microsoft Teams rely on ASR tools for subtitles or transcripts, errors in those outputs can intensify existing biases. 
Audiences may attribute transcription mistakes to the speaker rather than the system, further disadvantaging individuals with non-standard accents~\cite{kim2024same}.

Despite our results conflicting with previous work~\cite{kim2024same}, both studies suggest a path where ASR-generated subtitle errors harm both speaker and content evaluations and imply that the harms may disproportionately lie with speakers with non-standard accents. 
We encourage researchers and designers to explore strategies that mitigate these cascading harms. Improving model accuracy---such as through better training data from diverse dialects and speakers to reduce representational harms~\cite{cunningham-etal-2024-understanding}---is an important step. 
However, technical fixes alone are not enough.  
Addressing the biases that lead to unequal error rates will not entirely eliminate mistakes since ASR systems, like most machine-learning technologies, cannot attain complete accuracy, and external factors (e.g., environment noise or low-quality audio) can make the problem arbitrarily difficult.
Kim et. al~\cite{kim2024same}'s results suggest that identical ASR performance can be perceived differently depending on the speaker's accent, with subtitles judged as more helpful for speakers with non-standard accents than for those with standard accents. These findings point to an important, yet underexplored, dimension of fairness in ASR: \textit{perceived} subtitle quality.
While standard evaluation metrics such as word error rate treat all errors equally, users often weigh them differently (e.g., misrecognition of names or numbers may be more damaging than the omission of minor function words)~\cite{mishra2011predicting}. 
Perceived subtitle quality may therefore shape how audiences evaluate both the speakers and the content in ways that raw accuracy scores cannot capture.
The HCI community can help investigate how subtitle design and presentation (e.g., font size, placement, chunking, or timing) influence perceived quality and how adaptive or dynamic subtitle interfaces, those that change to maintain the same perceived subtitle quality based on the speaker, might mitigate harms caused by recognition errors. 
By integrating user-centered measures of perceived quality alongside traditional accuracy metrics, researchers and practitioners can help foster more equitable ASR systems and improve user experiences across diverse linguistic backgrounds.

Finally, we note that this work has some limitations that should be considered. First, our experiment relied on crowd participants who are likely Western, Industrialized, Educated,
Rich, and Democratic (WEIRD), and whose views (and perceptions) may differ from the US general public. Second, our experiment only manipulated the accent within a single demographic group (South Asian men). However, we have no reason to expect the results we observed are unique to this group. Prior research consistently shows that ASR systems across a wide range of languages exhibit systematically higher error rates for certain regional accents and dialects~\cite{Blaschke2025,gothi2023dialect}. These multilingual findings suggest that accent-related ASR disparities are a general phenomenon rather than one confined to a particular demographic group or language. Third, while we show the errors negatively impact the speaker and content evaluations, we did not examine the mechanism that causes the lower evaluations or determine whether the speaker or the system is blamed for the errors. Lastly, our experimental design relied on a between-subjects accent comparison, which reduced the statistical power to detect subtle accent effects. We deliberately selected a between-subjects design to avoid revealing the study purpose and contaminating participants' judgments.

\section{Conclusion}
Our findings show that subtitle quality significantly impacts speaker and content evaluations.
However, in this study, the speaker's accent did not affect the speaker evaluations or the content evaluations.
We encourage the research community to further explore subtitle quality and how it impacts speaker evaluations.
For example, what defines a good subtitle? While including additional information, such as coughs or filler words like ``umm'' and ``ahh'', may be factually correct, will it improve viewer perceptions? Moreover, which errors are significant? Audiences might be more forgiving of certain subtitle mistakes, while specific errors could directly harm a speaker's evaluation. 
These are important questions for future work.

\begin{acks}
A gift from the LinkedIn-Cornell Bowers CIS Strategic Partnership supported this research. Any opinions, findings, and conclusions or recommendations expressed in this material are those of the authors and do not necessarily reflect the views of LinkedIn.
We would also like to thank Danaé Metaxa, Allison Koenecke, and the members of the Social Technologies lab at Cornell Tech, whose feedback ultimately enriched this work.
\end{acks}

\bibliographystyle{ACM-Reference-Format}
\bibliography{chi26}

\appendix
\section{Appendix}

\subsection{Video Stimuli}
\label{appendix:vids}
\begin{figure}[h]
    \centering
    \begin{subfigure}[h]{0.45\textwidth}
        \centering
        \includegraphics[height=4.5cm]{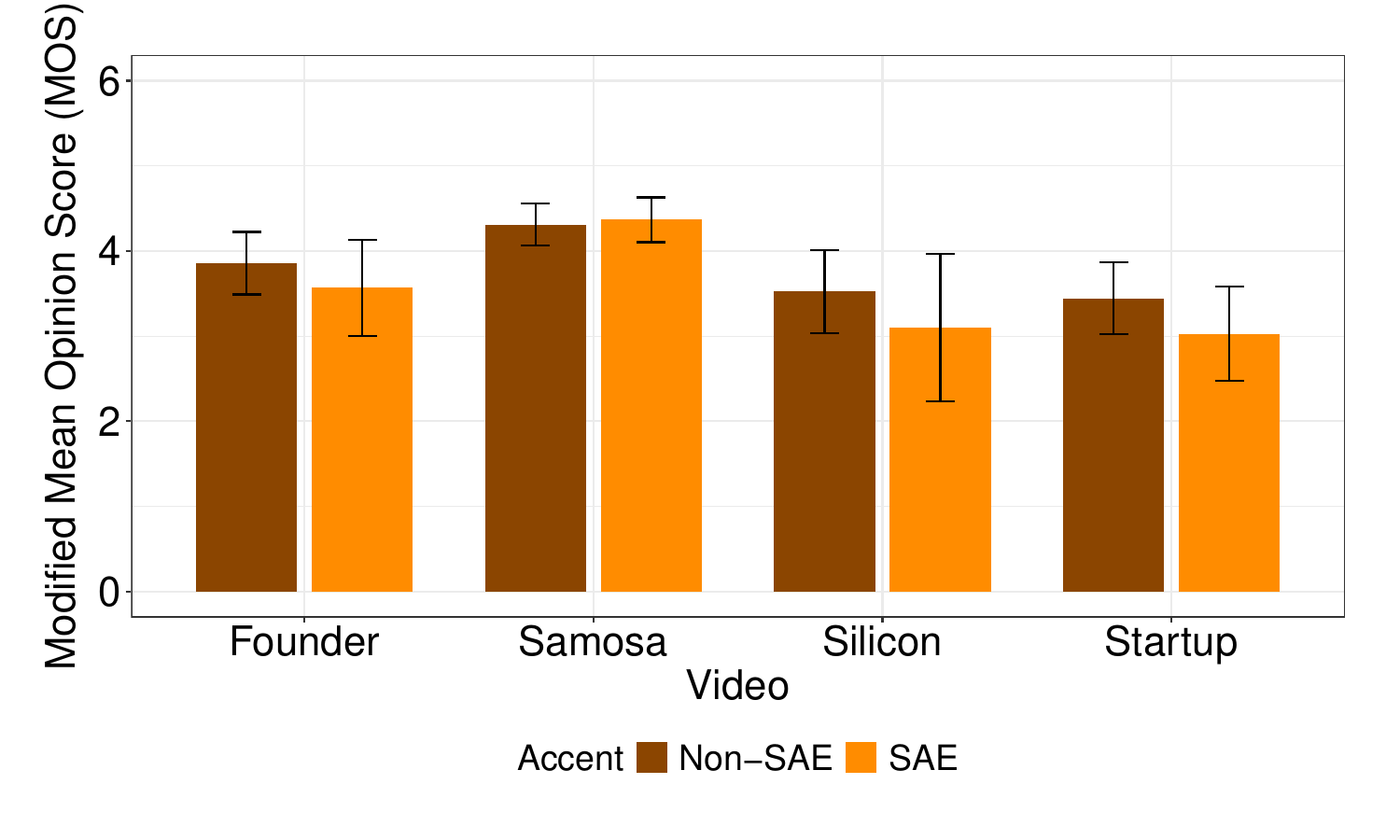}
        \caption{All of the audio sounded fairly natural.}
        \label{fig:voice_check}
    \end{subfigure} 
    \hspace{10pt}
    \begin{subfigure}[h]{0.45\textwidth}
        \centering
       \includegraphics[height=4.5cm]{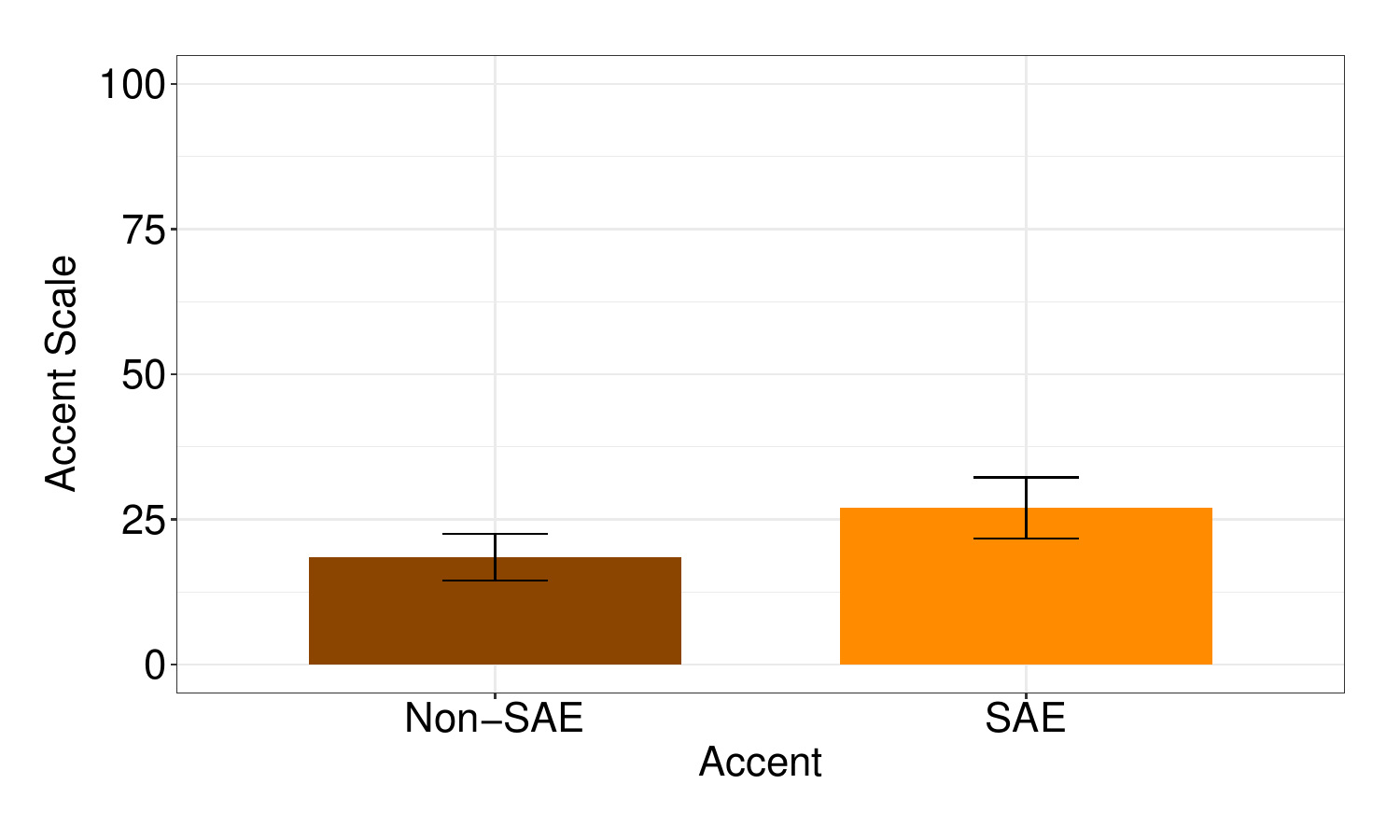}
       \caption{The non-SAE audio sounded less American than the SAE audio.}
       \label{fig:accent_check}
    \end{subfigure}
    \caption{\textbf{Audio Analysis.} All of the voices sounded natural, and there were subtle differences between the accent groups.}
    \label{fig:audio_analysis}
\end{figure}

We provide additional details on the video stimuli, beginning with the audio realism.
We recruited 30 participants from Prolific to listen to the audio snippets and label the quality of the voice using a modified Mean Opinion Score (MOS) measure~\cite{michel_voice_2025}. Participants were randomly placed into the SAE or non-SAE condition. After listening to each snippet, we asked participants to rate how natural the voice sounded, the flow of the audio, and whether the voice sounded robotic with a 5-point Likert scale. Figure~\ref{fig:voice_check} shows the mean and 95\% confidence intervals for the audio quality (Y-axis) for all videos (X-axis). Based on prior work, a score of 4.3-4.5 indicates excellent quality~\cite{michel_voice_2025}. The MOS for all of our videos is above 3, indicating a fair quality and greater. There were no significant differences in audio quality between the non-SAE audio and SAE audio.

We also confirm the difference between the accent groups with a between subjects pre-test experiment. Participants watched a 1-minute video clip of a speaker with a non-SAE or SAE accent. They were then asked to rank the accent on a scale from 0 (non-SAE) to 100 (SAE).
As seen in Figure~\ref{fig:accent_check}, the non-SAE accent ($M = 18.47, SD = 17.02$) was rated as less American than the SAE accent ($M = 26.93, SD = 22.50$).
A t-test confirmed the result to be significant ($p = 0.012$), although this accent difference did not produce a detectable effect in the main experiment.

\begin{figure}[h]
\includegraphics[width=\textwidth,height=4.5cm,keepaspectratio]{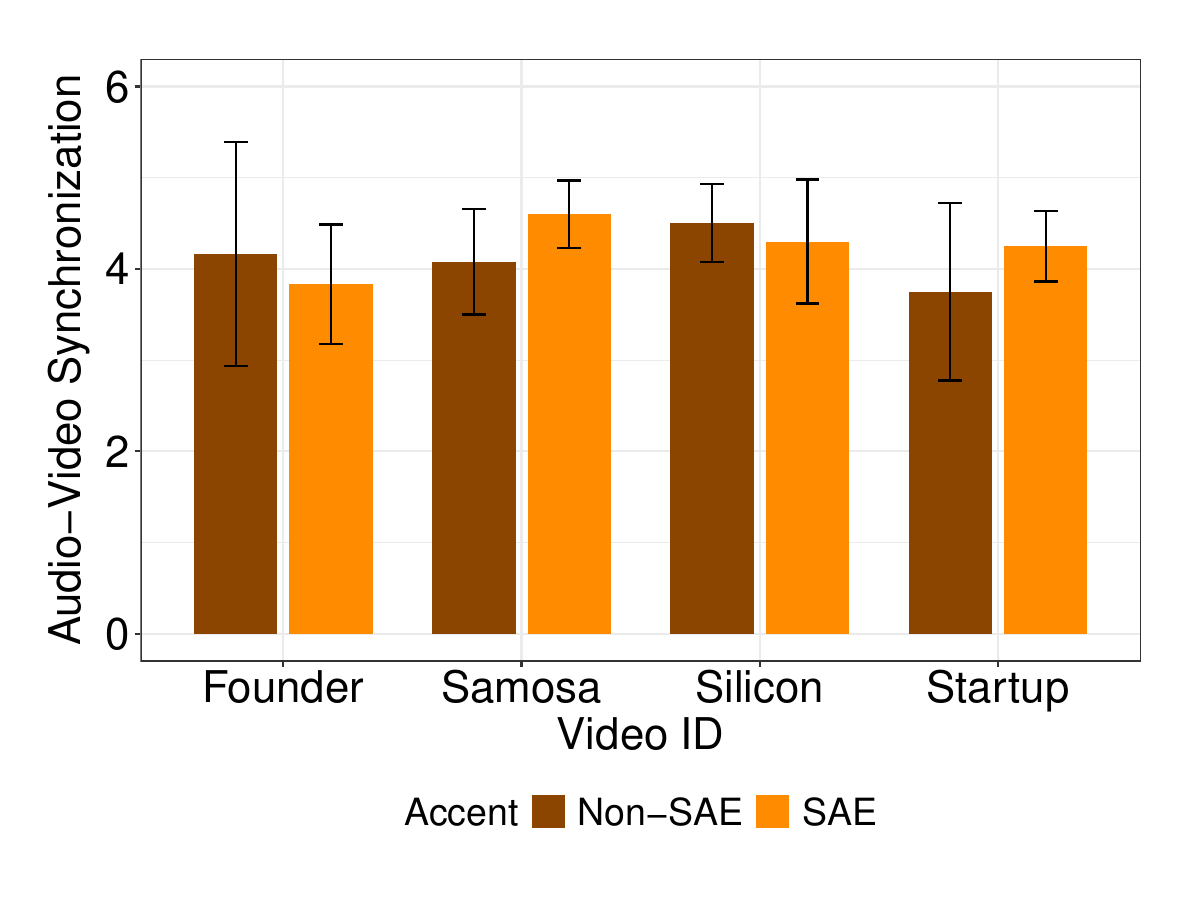}
\caption{\textbf{Audio Video Synchronization.} All of the videos had a medium synchronization.}
\label{fig:vid_check}
\end{figure}

Finally, we assessed audio–video synchronization using a separate between-subjects study in which participants watched the clips without subtitles to avoid distraction. After watching the video, we asked participants to rate the synchronization between the speaker's lip movements and the audio on a 5-point scale, based on the International Telecommunication Union's best practices for video quality. Figure~\ref{fig:vid_check} shows the mean and the 95\% confidence intervals for the synchronization. The means ranged from 3.75-4.6, indicating that there was a slight mismatch that did not interfere with the video watching experience. There were no significant differences in audio-video synchronization between the videos with non-SAE accents and SAE accents.

\subsection{Subtitles}
We provide additional insight into our subtitle development process with the word error rate of all our videos across different platforms in Table~\ref{tab:wer_results} and an example of our ground truth and error-prone subtitles in Table~\ref{table:subtitle_examples}.

\begin{table*}[h]
\centering
\caption{\textbf{ASR Performance.} Word error rates (WER) across transcription systems for different accents. Lower values indicate better performance.}
\label{tab:wer_results}
\begin{tabular}{lccccccc}
\toprule
\multicolumn{8}{c}{\textbf{Non-Standard American English Accents}} \\
\midrule
Video ID & Whisper & Google Meet & Zoom & Youtube & Otter & Apple & Adobe \\
\midrule
Samosa   & 0.2988 & 0.3402 & 0.1909 & 0.1909 & 0.2448 & 0.2324 & 0.2365 \\
Founder  & 0.1761 & 0.3920 & 0.2557 & 0.2386 & 0.2216 & 0.3409 & 0.1307 \\
Silicon  & 0.1312 & 0.3000 & 0.2563 & 0.1977 & 0.1813 & 0.2625 & 0.1250 \\
Start Up & 0.1173 & 0.1954 & 0.2123 & 0.1285 & 0.1397 & 0.1788 & 0.1844 \\
\midrule
Mean     & 0.1809 & \textbf{0.3070} & 0.2287 & 0.1879 & 0.1968 & 0.2534 & 0.1691 \\
\bottomrule
\end{tabular}

\vspace{0.5em} 

\begin{tabular}{lccccccc}
\toprule
\multicolumn{8}{c}{\textbf{Standard American English Accents}} \\
\midrule
Video ID & Whisper & Google Meet & Zoom & Youtube & Otter & Apple & Adobe \\
\midrule
Samosa   & 0.1452 & 0.3859 & 0.2033 & 0.1618 & 0.1992 & 0.2863 & 0.2158 \\
Founder  & 0.1875 & 0.3864 & 0.3011 & 0.2784 & 0.2443 & 0.3977 & 0.1705 \\
Silicon  & 0.2375 & 0.2750 & 0.2063 & 0.1813 & 0.1938 & 0.2438 & 0.1313 \\
Start Up & 0.1341 & 0.2061 & 0.2234 & 0.1117 & 0.1508 & 0.2011 & 0.1955 \\
\midrule
Mean     & 0.1761 & \textbf{0.3135} & 0.2335 & 0.1833 & 0.1970 & 0.2823 & 0.1783 \\
\bottomrule
\end{tabular}
\end{table*}

\begin{table*}[h]
\caption{\textbf{Transcription Errors}. We provide a sample of the ground truth and the error-prone transcriptions for a sample of our videos. Noticeable differences in the sentences are highlighted in blue.}
\centering
\begin{tabular}{|p{8cm}|p{8cm}|} 
\hline
\textbf{Ground Truth} & \textbf{Transcription} \\ \hline
Cool year for us. But every single one of these guys asked me, Munaf, what's your vision with the Borhi Kitchen? And TBK pretty much is a journey...  \newline \newline

I went into a wealth management role, nice cushy job. I was enjoying my life. \newline \newline

And he just told me what calls you? And he said Manav just be available and something will call you, and this called me. So I left my corporate job   
&  
Coolio for us, but every single one of these guys asked me Munaf, what's your vision with the  \textcolor{blue}{body} kitchen? And  \textcolor{blue}{TV came} pretty much is a journey...\newline\newline

I went into  \textcolor{blue}{[omitted 'a']} wealth management role. Nice  \textcolor{blue}{khushi} job. I was enjoying my life.\newline\newline 

And he just told me what calls you \textcolor{blue}{[omitted '?']} and he said,  \textcolor{blue}{Man of} just be available. And something will call you and this called me. So I  \textcolor{blue}{met my pocket} job.  
\\ \hline
\end{tabular}
\label{table:subtitle_examples}
\end{table*}

\subsection{Participant Demographics}
\label{appendix:dems}
We recruited a U.S.-based population on Prolific and had usable data from 207 participants. Table~\ref{table:demographics} provides the full details on the participants in our final dataset.

\begin{table*}[h]
\caption{\textbf{Participants' Demographic Background}. An overview of participants' demographic background.}
\centering
\begin{tabular}{|l r|}
\hline
\multicolumn{1}{|c|}{\textbf{Category}} & \multicolumn{1}{c|}{\textbf{Percentage}} \\ \hline
\multicolumn{2}{|l|}{\textbf{Age}} \\ \hline
18–24 years old & 5.31\% \\ 
25–34 years old & 14.49\% \\ 
35–44 years old & 24.64\% \\ 
45–54 years old & 33.33\% \\ 
55–64 years old & 18.84\% \\ 
65+ years old & 3.38\% \\ \hline
\multicolumn{2}{|l|}{\textbf{Gender}} \\ \hline
Female & 53.14\% \\ 
Male & 45.89\% \\ 
Non-binary / third gender & 0.97\% \\ \hline
\multicolumn{2}{|l|}{\textbf{Education}} \\ \hline
Less than high school degree & 0.97\% \\ 
High school graduate (or equivalent) & 11.59\% \\ 
Some college but no degree & 15.94\% \\ 
Associate degree (2-year) & 10.63\% \\ 
Bachelor’s degree (4-year) & 40.58\% \\ 
Master’s degree & 16.43\% \\ 
Professional degree (JD, MD) & 1.45\% \\ 
Doctoral degree & 2.42\% \\ \hline
\multicolumn{2}{|l|}{\textbf{Race}} \\ \hline
Asian & 3.38\% \\ 
Black / African American & 15.94\% \\ 
Hispanic & 2.42\% \\ 
Prefer to self-describe & 1.45\% \\ 
White / Caucasian & 76.81\% \\ \hline
\end{tabular}
\label{table:demographics}
\end{table*}

\end{document}